\documentclass{sig-alternate}
\usepackage{array}
\usepackage{caption}
\usepackage{cite}
\usepackage[usenames,dvipsnames]{color}
\usepackage{url}
\newcommand{\subparagraph}{}
\usepackage[compact]{titlesec}
\usepackage{mdwlist}

\titlespacing{\section}{0pt}{4pt}{*0}
\titlespacing{\subsection}{0pt}{5pt}{*0}
\titlespacing{\subsubsection}{0pt}{*1}{*0}
\begin{document}
\conferenceinfo{SIGIR 2014}{Gold Coast, Australia}

\title{Interactions between \protect\\Health Searchers and Search Engines}

\numberofauthors{2} 

\author{
\alignauthor
George Philipp\titlenote{Research done during an internship at Microsoft Research.}\\
       \affaddr{Carnegie Mellon University}\\
       \affaddr{5000 Forbes Avenue}\\
       \affaddr{Pittsburgh, PA 15213}\\
       \email{george.philipp@email.de}
\alignauthor
Ryen W. White\\
       \affaddr{Microsoft Research}\\
       \affaddr{One Microsoft Way}\\
       \affaddr{Redmond, WA 98052}\\
      \email{ryenw@microsoft.com}
}

\maketitle

\begin{abstract}
The Web is an important resource for understanding and diagnosing medical conditions. Based on exposure to online content, people may develop undue health concerns, believing that common and benign symptoms are explained by serious illnesses. In this paper, we investigate potential strategies to mine queries and searcher histories for clues that could help search engines choose the most appropriate information to present in response to exploratory medical queries. To do this, we performed a longitudinal study of health search behavior using the logs of a popular search engine. We found that query variations which might appear innocuous (e.g. ``bad headache'' vs ``severe headache'') may hold valuable information about the searcher which could be used by search engines to improve performance. Furthermore, we investigated how medically concerned users respond differently to search engine result pages (SERPs) and find that their disposition for clicking on concerning pages is pronounced, potentially leading to a self-reinforcement of concern. Finally, we studied to which degree variations in the SERP impact future search and real-world health-seeking behavior and obtained some surprising results (e.g., viewing concerning pages may lead to a short-term reduction of real-world health seeking).

\end{abstract}

\setlength{\parindent}{0pt} 
\category{H.3.3}{Information Storage and Retrieval}{Information Search and Retrieval}[Search process, Query formulation]
\keywords{Health search, medical search,  diagnosis, log/behavioral analysis, cyberchondria}

\setlength{\parindent}{9pt} 
\section{INTRODUCTION}

Health anxiety is a significant problem in our modern medical system \cite{asmundson, taylor}. The belief that one's common and benign symptoms are explained by serious conditions may have several adverse effects such as quality-of-life reduction, incorrect medical treatment and inefficient allocation of medical resources. The web has been shown to be a significant factor in fostering such attitudes \cite{whitehorvitzcyber, whitehorvitzhypo}. A recent study found that 35\% of U.S. adults had used the Web to perform diagnosis of medical conditions either for themselves or on behalf of another person, and many ($>$50\%) pursued professional medical attention concerning their online diagnosis \cite{foxduggan}. Motivated by the popularity of online health search, we investigated how search engines might improve their health information offerings. We hypothesize that searchers are less likely to develop unrealistic beliefs when they are given unbiased and well-presented information about their medical state. Thus, we believe an ideal search engine should use queries, and available search histories, to extract medically relevant information about the individual as well as detect any health anxieties.

Let us give a (negative) example of a possible medical search session. A user experiencing anxiety about a headache might first spend some time searching for information on a serious condition such as ``brain tumor'' and then switch to a symptom query about headache (a type of transition that prior work shows occurs frequently \cite{cartright}). The user might choose the query wording ``severe headache explanations'' because of the subjective concern they experience at query time. The engine, registering the words ``severe'' and ``explanations'' as well as the phrase ``brain tumor" present in the user's search history might compile a search engine result page (SERP) that is biased towards serious conditions. The user, viewing the SERP through the lens of their current health anxiety, may be attracted towards serious conditions in captions \cite{whitehorvitzbias} and hence select a concerning page, heightening their anxiety further.

In this paper, we highlight a range of challenges and opportunities in working towards improving exploratory health search and thus hope to outline an agenda that frames this problem. Achieving this requires an understanding of both user and search engine behavior. Users play a role in the search process in two ways: their choice of query formulation as well their subjective consumption of information on the SERP and pages clicked on. This naturally led us to formulate three research questions:
\begin{itemize*}
\item [{\bf Q1}] How does a user's subjective medical concern shape his or her choice of wording for medical queries?
\item [{\bf Q2}] How does the search engine interpret the subjective medical concern and objective medical information expressed in the query as well as other measurable characteristics (such as medical search history) when compiling the SERP?
\item [{\bf Q3}] How do users respond to the SERP, both in terms of the consumption of the information on the SERP as well as changes in future behavior caused by viewing the SERP and pages clicked on?
\end{itemize*}

Results pertaining to these questions are found in Sections 4, 5, and 6 respectively. To answer them, we studied the search logs of 190 thousand consenting users of a major commercial Web search engine. Search logs are a valuable resource for studying information seeking in a naturalistic setting, and such data has been used by several studies to explore how searchers obtain medical information \cite{whitehorvitzcyber, cartright}. The pipeline used to process this data and the features extracted for analysis are further described in Section 3. 

The main contributions from our analysis are:

\begin{itemize*}
\item Revealing how certain users have specific preferences for certain query formulations (e.g. ``i have muscle pain'' vs. ``muscle pain'') which also has a significant effect on search results, and potentially health outcomes. 
\item Finding evidence that users might not be swayed by concerning content appearing on SERPs as we might expect based on prior studies.
\item Quantifying the extent to which users with prior medical concern receive more concerning SERPs in response to health queries and choose the most concerning pages to click on, potentially leading to a vicious cycle of concern reinforcement. 
\item Determining to how users directly and indirectly influence the level of concern expressed in the SERP they receive, both through query choice and other factors (e.g., personalization).
\end{itemize*}

We discuss these findings and their implications in Section 8 and conclude in Section 9.

\section{RELATED WORK}

Related research in this area falls into three main categories: health search behavior; the quality of online health content; and health anxiety and the impact of reviewing health content on the Web.

There continues to be interest in search and retrieval studies on expert and consumer populations in a variety of domains, typically conducted as laboratory studies of search behavior \cite{bhavnani, hershhickam}. Benigeri and Pluye \cite{benigeripluye} showed that exposing novices to complex medical terminology puts them at risk of harm from self-diagnosis and self-treatment. It is such consumer searching (rather than expert searching) that we focus on in the remainder of this section.

Search engine log data can complement laboratory studies, allowing search behavior to be analyzed at scale in a naturalistic setting and mined for a variety of purposes. Logs have been used to study how people search \cite{jansen}, predict future search and browsing activity \cite{lauhorvitz}, model future interests \cite{dupretpiwowarski}, improve search engine quality \cite{joachims}, and learn about the world \cite{richardson}. Focusing on how people perform exploratory health searching, Cartright et al. \cite{cartright} studied differences in search behaviors associated with diagnosis versus more general health-related information seeking. Ayers and Kronenfeld \cite{ayerskronenfeld} explored changes in health behavior associated with Web usage, and found a positive correlation between it and the likelihood that a user will change their health behavior based on the content viewed. 

The reliability of the information in search results is important in our study; unreliable information can drive anxiety. The quality of online healthcare information has been subject to recent scrutiny. Lewis \cite{lewis} discussed the trend toward accessing information about health matters online and showed that young people are often skeptical consumers of Web-based health content. Eysenbach and Kohler \cite{eysenbachkohler} studied users engaged in assigned Web search tasks. They found that the credibility of Web sites was important in the focus group setting, but that in practice, participants largely ignored the information source. Sillence and colleagues \cite{sillence} studied the influence of design and content on the trust and mistrust of health sites. They found that aspects of design engendered mistrust, whereas the credibility of information and personalization of content engendered trust. 

The medical community has studied the effects of health anxiety, including hypochondriasis \cite{asmundson}, but not in Web search. Health anxiety is often maladaptive (i.e., out of proportion with the degree of medical risk) and amplified by a lack of attention to the source of their medical information \cite{taylor, kring}. Such anxiety usually persists even after an evaluation by a physician and reassurance that concerns lack medical basis. A recent study showed that those whom self-identified as hypochondriacs searched more often for health information than average Web searchers \cite{whitehorvitzhypo}. By estimating the level of health concern via long-term modeling of online behavior, search engines can better account for the effect that results may have and help mitigate health concerns. Our research makes progress in this area.

\begin{table*}
\centering
\begin{tabular}{>{\centering\arraybackslash}p{\textwidth}} 
headache, headaches, severe headache what do I do, which remedy for headache, headache top of head with back pain \\ \hline
headache do I have a {\it tumor}, {\it headache rack}, my {\it job} gives me a headache, headache {\it national parks of california} \\ \hline 
a, low, be, he, sick, she, helps, standing, black, speech, male, between, acute, shaking, sensitive, bending, an, testing \\ \hline
\end{tabular}
\captionsetup{justification=centering}
\caption{Examples of landmark queries (top), of non-landmark queries (middle; secondary topics and medical conditions are italic) and of admissible words we used to find potential landmark queries (bottom)}
\end{table*}

Searchers may feel too overwhelmed by the information online to make an informed decision about their care\cite{hart}. Performing self-diagnosis using search engines may expose users to potentially alarming content that can unduly raise their levels of health concern. White and Horvitz \cite{whitehorvitzcyber} employed a log-based methodology to study escalations in medical concerns, a behavior they termed cyberchondria. Their work highlighted the potential influence of several biases of judgment demonstrated by people and search engines themselves, including base-rate neglect and availability. In a follow up study \cite{whitehorvitzesc}, the same authors showed a link between the nature and structure of Web page content and the likelihood that users' concerns would escalate. They built a classifier to predict  escalations associated with the review of content on Web pages (and we obtained that classifier for the research described in this paper). Others have also examined the effect of health search on user's affective state, showing that the frequency and placement of serious illnesses in captions for symptom searches increases the likelihood of negative emotional outcomes \cite{lauckner}. Other research has shown that health-related Web usage has been linked with increased depression \cite{bessiere}.

Moving beyond the psychological impact of health search, researchers have also explored the connection between health concerns and healthcare utilization. In one study \cite{whitehorvitzhui}, the authors estimated that users sought medical attention by identifying queries containing healthcare utilization intentions (HUIs) (e.g., [physicians in san jose 95113]). Eastin and Guinsler \cite{eastinguinsler} showed that health anxiety moderated the relationship between health seeking and healthcare utilization. Baker and colleagues \cite{baker} examined the prevalence of Web use for healthcare, and found that the influence of the Web on the utilization of healthcare is uncertain. The role of the Web in informing decisions about professional treatment needs to be better understood. One of our contributions in this paper is to demonstrate the potential effect of health-related result pages on future healthcare utilization.

Our research extends previous work in a number of ways. By focusing on the first query pertaining to a particular symptom observed in a user history, we show that small differences in query formulation can reflect significant differences amongst health searchers and their health-related search outcomes. To date, no research has demonstrated the impact and insight afforded from analyzing such landmark queries and the behavior around them. To our knowledge, we are also the first to use the search logs to devise statistical experiments which allow us to quantify effects such as user response to medical search results amongst real user populations and provide evidence for causal relationships where possible. We believe that such an approach is necessary to formulate definitive implications for search engine design as well as measuring search engine performance. 

\section{STUDY}
We describe various aspects of the study that we performed, including the data, the features extracted, and the statistical methods used for analysis.

\subsection{Log Data}

To perform our study we used the anonymized search logs of a popular search engine. Users of this search engine give consent to having information about their queries stored via the terms of use of the engine. During this study, we focused on medical queries related to headache, as it is among the most common health concerns \cite{nielsen}. We use the phrase {\it headache query} to refer to queries that contain the substring ``headache'' and occurred during the six month period from September 2012 to February 2013. Amongst those, we call a {\it landmark query} a query that shows an intent to explore the symptom ``headache'', that does not already contain a possible explanation for headache (e.g., migraine, tumor) and that is not otherwise off-topic. We found these landmark queries by manually assessing frequent headache queries and creating a list of 682 ``admissible'' terms that we believed could occur in landmark queries. We then compiled all queries that exclusively contain terms from that list into a dataset. Examples of landmark queries, non-landmark queries, and admissible words can be found in Table 1. From manual inspection, we concluded with confidence that the dataset captured over 50\% of all landmark queries present in the logs and that over 95\% of captured queries are proper landmark queries, the rest being headache queries which contain a significant secondary topic. We then excluded all but the first landmark query instance for each user, ensuring that each user only appears once in the dataset. Overall, our dataset contains over 50,000 unique queries and over 190,000 query instances / users.

We focus on headache since it a common medical symptom (e.g., over 95\% of adults report experiencing headaches in their lifetime \cite{rasmussen}) and there are a variety of serious and benign explanations (from caffeine withdrawal to cerebral aneurism), facilitating a rich analysis of content, behavior, and concern. While we believe that headache searching is sufficiently rich and frequent to warrant its own study, investigating queries related to symptoms other than headache could solidify our findings. A large-scale analysis similar to that reported here, but focused on multiple symptoms, is an interesting and important area for future work.

\subsection{Features}

For each landmark query in our dataset, we generated features. To frame our three research questions, we modeled the search process around a landmark query as five separate stages: (i) The user's search behavior prior to the landmark query, (ii) the user's choice of wording of the landmark query, (iii) the SERP returned to the user by the search engine, (iv) the user's decision which pages to click on (if any), and (v) the user's search behavior after the landmark query. Our research questions ask about the relationship of these 5 stages. Hence, the features we extracted come in five groups.

\begin{table}
\centering
\begin{tabular}{lm{2.3in}cc} 
Name&Description  \\ \hline
pasttopserious&Number of medical queries containing the most frequent serious condition amongst all serious conditions present \\ \hline
pasttopbenign&Number of medical queries containing the most frequent benign condition amongst all benign conditions present \\ \hline
pastdiffserious&Number of distinct serious conditions in medical queries \\ \hline
pastdiffbenign&Number of distinct benign conditions in medical queries\\ \hline
pastmedical&Number of medical queries \\ \hline
pastheadache&Number of medical queries containing ``headache'' or a condition that is an explanation for headache \\ \hline
pasthui&Number of HUI queries \\ \hline
pasthuiclicked&Number of HUI queries where at least one page on the SERP was clicked \\ \hline
\end{tabular}
\caption{User Search Behavior Features}
\end{table}

\begin{table*}
\centering
\begin{tabular}{lllc} 
Name & Description (query ..) & Example & Frequency\\ \hline
audience & .. is about specific population &``headache in adults'' & 3.6\%\\ 
filler & .. contains a filler such as 'a' or 'and' &``headache and cough'' & 3.6\%\\ 
goal & .. contains a specific search goal & ``definition of headache'' & 26.3\%\\ 
goal:condition & .. indicates the goal of diagnosis & ``reasons for headache'' & 8.2\%\\ 
goal:symptom & .. indicates the goal of related symptoms & ``headache symptoms'' & 2.1\%\\ 
goal:treatment & .. indicates the goal of treatment & ``headache cure'' & 14.7\%\\ 
goal:medication & .. indicates the goal of treatment through medication & ``headache pills'' & 2.4\%\\ 
goal:alternative & .. indicates the goal of alternative treatment & ``natural headache remedy'' & 5.4\%\\ 
symptomhypothesis & .. states another symptom as cause & ``headache caused by back pain'' & 1.7\%\\ 
eventhypothesis & .. relates the headache to a life event & ``headache after hitting head'' & 3.3\%\\ 
duration & .. specifies a duration for the headache & ``chronic headache'' & 10.9\%\\ 
intensity & .. specifies that the headache is strong & ``severe headache'' & 5.0\%\\ 
location & .. specifies the location of the headache & ``headache left side of head'' & 17.3\%\\ 
pronoun & .. contains a pronoun & ``i have headache'' & 5.8\%\\ 
kindofheadache & .. specifies the kind of pain & ``stabbing headache'' & 2.5\%\\ 
othersymptom & .. specifies an additional symptom & ``headache reflux'' & 27.9\%\\ 
triggered & .. indicates a headache trigger & ``headache when bending over'' & 1.4\%\\ 
timeofday & .. indicates a daily pattern & ``headache in afternoon'' & 3.2\%\\ 
openquestion & .. is phrased as an open question & ``what to do about headache'' & 11.1\%\\ 
\end{tabular}
\caption{High-level {\it QueryFormulation} features. Note that queries may have multiple features, such as ``severe headache on top of head and cough'', so the percentages in the far-right column do not sum to 100\%.}
\end{table*}

{\it BeforeSearching}. This group of features describes the level of medical searching before the landmark query (stage (i)). For each query in the user's search log, we first extracted whether that query was of a medical nature. For this, we used a proprietary classifier. From manual inspection, we concluded with confidence that its Type I and II errors are $< 0.1$. Queries so classified as medical will be called {\it medical queries}. Secondly, we extracted phrases present in the query that describe medical conditions, such as ``common cold'' or ``cerebral aneurism''. The list of phrases we considered was based on the International Classification of Diseases 10th Edition (ICD-10) published by the World Health Organization as well as the U.S. National Library of Medicine's PubMed service and other Web-based medical resources. We also used manually curated synonyms from online dictionaries and standard grammatical inflections to increase coverage. For more information see the approach used by \cite{whitehorvitzcyber}. The list was also separated into benign and serious medical conditions and we determined which conditions are possible explanations for headache. Thirdly, we extracted phrases that indicate the query is linked to an intention of real-world healthcare utilization (HUI), such as ``emergency care'' or ``hepatologist''. We call those queries {\it HUI queries}. Finally, the individual-query-level features were aggregated over a time window just before the landmark query to form 8 distinct features, which are shown in Table 2. In our experiments, we considered five different aggregation windows: 1 hour, 1 day, 1 week, 30 days and 90 days. All experiments involving {\it BeforeSearching} features were replicated for each aggregation window. 

We believe that intense medical searching may be a sign of health concerns or anxieties. White and Horvitz \cite{whitehorvitzhypo} extensively demonstrated that users believing their symptoms may be explained by a serious condition conduct longer medical search sessions and do so more frequently. Of course, there are many potential reasons for increased medical search such as different web search habits, random noise or even a recent visit to a physician \cite{whitehorvitzhui}. Overall, we believe that it makes sense to view {\it BeforeSearching} in light of possible medical concern experienced. However, even if there are significant other factors at play, we believe that the phenomenon of health search intensity remains interesting.

{\it QueryFormulation}. The choice of wording for the query (stage (ii)) was modeled, firstly, using 19 high-level features. We arrived at these by manually inspecting admissible words and landmark queries (such as in Table 1) and noting the most important high-level ideas expressed through them. We believe that those 19 features capture a significant portion of the semantic variation within queries. The features are shown in Table 3. They offer a useful characterization of the broad range of different types of search intent associated with headache-related queries. Four of these features ({\it othersymptom}, {\it duration}, {\it intensity} and {\it location}) were further divided by which key phrase was used to express this feature, yielding an additional 117 low-level features (examples are shown in the graph annotation of Figure 1.2-1.5). All feature extraction functions are based on substring matches joined by logical operators. From manual inspection, we concluded with confidence that the Type I and II errors of the extractors of all these features was low. All features in this category are binary.

\begin{figure*}
\centering
\includegraphics[scale=1.8]{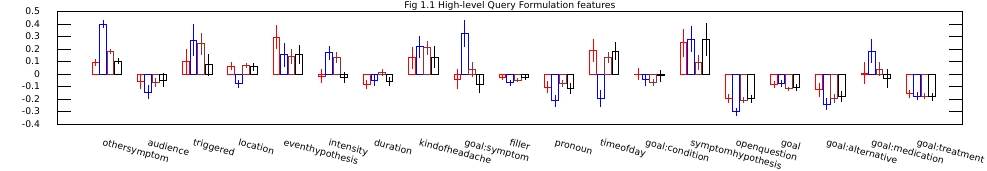}
\includegraphics[scale=1.8]{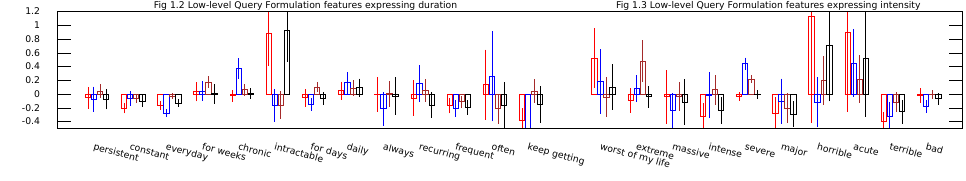}
\includegraphics[scale=1.8]{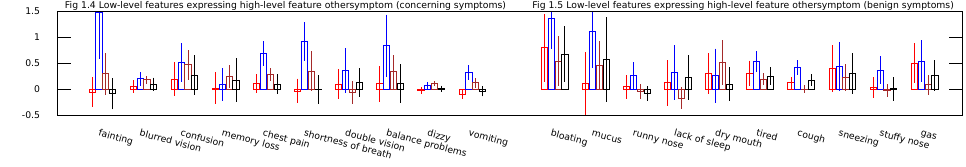}
\caption{Association between query formulation and search behavior features. From left to right: {\bf {\color{red} Red}}: {\it pastmedical} (aggregation window: 90 days); {\bf {\color{blue} Blue}}: {\it pastmedical} (window: 1 hour); {\bf {\color{Brown} Brown}}: {\it futuremedical} (window: 1 hour); {\bf Black}: {\it futuremedical} (window: 90 days). All bars show the relative change of medical search intensity of users whose landmark query has a certain formulation relative to the global mean. Error bars are 95\% confidence intervals.}
\end{figure*}

{\it SERPConcern}. We scored the level of medical concern expressed in each of the Top 8 pages on the SERP (stage (iii)) using a logistic regression classifier designed to predict searching for serious conditions and shown in \cite{whitehorvitzesc} to have significant predictive power. The classifier was graciously provided to us by the authors for use in our study. Note that during the time period analyzed in this study, the search engine only returned eight results on a large number of SERPs so we disregarded possible further results. It has been shown that users rarely click below position 8, including for health queries \cite{whitehorvitzbias}. The classifier is based on page features from the URL and HTML content similar to those shown in Table 2. It also includes features that attempt to measure the ``respectability'' of the page (e.g., expressed through the {\it Health on the Net Foundation} certificate \cite{hon}) to capture the effect of this on the user. It is then trained to discriminate between pages that lead to serious condition searches (concerning) and pages that lead to benign condition searches (non-concerning). The concern score of an arbitrary page is then the inner product between its feature vector and the learnt weight vector. Finally, we take the weighted sum of the 8 scores for the individual pages on any SERP to produce a single feature value for the full SERP. Note that even though we consider pages leading to benign condition searches less concerning than those leading to serious condition searches, throughout this study, we still consider benign condition searching as an indicator for medical concern experienced by the user, albeit less strong than serious condition searching.

{\it ClickFeatures}.  To measure the users click behavior (stage (iv)), we record whether the user clicked a page on the SERP. If the user did, we also recorded the concern score of the page clicked (as described in the last paragraph) as well as the position of the clicked page on the SERP. We call these three features {\it hasclicked}, {\it clickconcern} and {\it clickposition}. (Users that did not click are excluded from analysis involving features {\it clickconcern} and {\it clickposition}.)

{\it AfterSearching}. The same 8 features as {\it BeforeSearching}, but aggregated over a window just after the landmark query (stage (v)). In the name, we replace ``past'' with ``future'' (e.g. one feature is called {\it futuremedical}).

\subsection{Statistical Methodology}

Let $X$ be a dataset where each entry corresponds to a user / landmark query (either our full dataset of 190.000 entries or a subset of it.). Write $x_1, .., x_N$ for the data points and $x^1_n, .., x^d_n$ for the components / feature values of data point $x_n$. Throughout our analysis, we wish to measure whether there is a significant association between two feature values $i$ and $j$, say between {\it pastmedical} and {\it SERPConcern}. By this we mean that the p-value of a suitable independence test on the two variables is low. To measure this simply and robustly, we will choose a threshold $t^i$ and split the data set into two subpopulations $X_<$ and $X_>$ such that for all $x_n$, $n \in \{1, .., N\}$, we have $x_n \in X_< \iff x_n^i < t^i$. We also call the two subpopulations the {\it lower bucket} and {\it upper bucket} respectively. Then, we either perform a two-sample t-test on the population means of $X_<$ and $X_>$ (Figure 3.3) or we consider a 95\%-confidence interval around the mean of the smaller bucket if it is significantly smaller than the other bucket (Figures 1 and 2). (In practice, all our subpopulations are large enough to warrant the use of gaussian confidence intervals / tests.) A statistical association between features $i$ and $j$ is a symmetric relation, we may choose to split on either feature and compare the means of the other, based on convenience of presentation.

Several times, we will encounter a more challenging case where we want to answer the question whether two feature values $i$ and $j$ are associated while controlling for a third feature value $k$. We do this by first dividing the dataset into many subpopulations according to the exact value of feature $k$ to obtain $X^{v^k_1}$, .., $X^{v^k_{N^k}}$, where $v^k_1, .., v^k_{N^k}$ are the values $x_n^k$ can take. Then, we split each of these subpopulations further according to thresholds on feature $i$ as before. Hence, we obtain $X_<^{v^k_1}$, .., $X_<^{v^k_{N^k}}$ and $X_>^{v^k_1}$, .., $X_>^{v^k_{N^k}}$. Because it is nontrivial to jointly compare this potentially large number of buckets, we adopt the following 2-step procedure. 

First, we take the union of all lower and upper buckets respectively and perform a two-sample t-test as before. For this to control for feature $k$, each lower bucket must be of the same size as its corresponding upper bucket. If, for some $X^{v^k_n}$ there is no threshold that achieves this, we randomly remove data points whose feature value is equal to the median across $X^{v^k_n}$ until this is possible. While this significantly reduces the number of data points entering the analysis, we do not believe this threatens validity or generalizability.

In the second step, we first individually compare each lower bucket to its corresponding upper bucket. Because many of these buckets are small (e.g. of size 1), we use the Mann-Whitney U statistic to obtain a p-value for each pair of buckets. Then, we aggregate all of those p-values by Stouffer's Z-score method where each p-value is weighted according to the size of its respective buckets. 

Finally, we wish to combine p-values obtained in both steps to either accept or reject the null-hypothesis at a given significance level. This is achieved by taking the minimum of both values and multiplying by 2. The cumulative distribution function of that combined quantity is below that of the uniform distribution under the null hypothesis and is thus at least as conservative in rejecting the null as each individual p-value, while gaining a lot of the statistical power of both tests. We quote this value in Figure 3.1-3.2 and 3.4-3.7.

\section{QUERY FORMULATION}

\begin{figure*}
\centering
\includegraphics[scale=1.8]{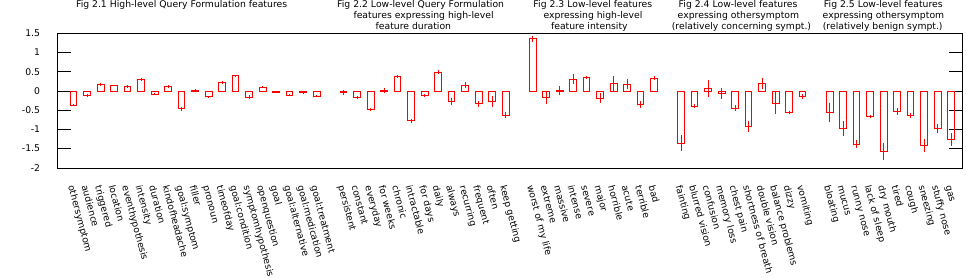}
\caption{Association between query formulation features and {\it SERPConcern}. All bars show the difference in {\it SERPConcern} of users whose landmark query has a certain formulation relative to the global mean, measured in standard deviations of the global distribution.}
\end{figure*}

We began our analysis by investigating how users with different levels of {\it BeforeSearching} phrase their queries. Figure 1.1 shows the mean value of {\it pastmedical} for each high-level {\it QueryFormulation} feature relative to the population mean, both aggregated over 1 hour and 90 days. Error bars indicate 95\% confidence intervals. We chose the feature {\it pastmedical} to represent {\it BeforeSearching} because it is the least sparse and hence has the smallest confidence interval.

To our surprise, we found that most {\it QueryFormulation} features are highly discriminative, i.e. users choosing queries with certain features have conducted significantly more medical searching than users choosing queries that have different features. In fact, every one of these 19 features is associated with a significant change in prior medical search activity during the hour before the landmark query. This effect could be explained by users near the beginning of their medical search process choosing different query patterns compared to users near the end of their medical search process. However, most {\it QueryFormulation} features are still discriminative when aggregated over a 90 day window (red bars), and we found that the majority of medical searches in that window do not occur immediately prior to the landmark query. Figure 1.1 also shows the feature {\it futuremedical}. We find that most {\it QueryFormulation} features have the same association with past and future search activity, even over a 90 day window. This is evidence that these {\it QueryFormulation} features characterize users and that heavy medical searchers prefer certain formulations compared to other users in a consistent fashion over the long term.
 
Over the 1 hour window, users entering an additional symptom in their landmark query show the highest level of search activity. This might be because experiencing a larger number of symptoms causes the user to want to find information about each individual symptom, thus increasing the search need. Surprisingly, {\it intensity} is not one of the strongest predictors of increased searching even though it appears to be the most intuitive indicator of increased concern. Also, the features {\it openquestion} and {\it pronoun} indicate less prior searching. Hence, the presence of sentence-like structures in queries is potentially associated with lower user health concern. It is difficult to intuitively interpret most of the {\it QueryFormulation} features. Figure 1.2-1.5 shows low-level {\it QueryFormulation} features relating to high-level features {\it duration}, {\it intensity} and {\it othersymptom}. Unfortunately, the sparseness of {\it pastmedical} leads to large margins of error, the exact size of which are also difficult to determine. Nonetheless, certain features which are more common and thus have lower margins or error such as ``constant'' or ``terrible'' are as discriminative as high-level features, suggesting they are useful for making inferences about the user. For example, our analysis suggests the very counterintuitive conclusion that users searching for ``headache everyday'' experience a different level of concern than users searching for ``daily headache'' (see difference in feature {\it pastmedical} between {\it daily} and {\it everyday} (blue bar) in Figure 1.2). Similarly, different symptoms are associated with their own level of past search activity. We chose to display 10 relatively concerning symptoms (Figure 1.4) and 10 relatively benign symptoms (Figure 1.5). The choice was made using the findings of a separate crowdsourcing study that we omit from the paper due to space reasons. In that study, many participants were asked to estimate the level of medical concern associated with a set of symptoms. Even though we do not present this study, the difference between the two groups of symptoms is intuitively clear. Surprisingly, we do not see a clear trend that users under Figure 1.4 have searched more, which weakens the hypothesis that objective medical state is linked to search activity.

In summary, we find that both high-level {\it QueryFormulation} features as well as individual word choices reveal information about the searcher, which is not necessarily expected. While some {\it QueryFormulation} features are interpretable, more work is necessary to understand their precise meaning. Also, more data is needed to better study their effect on rarer search events such as HUI queries. 

\section{EFFECT ON SERP}

When personalization is employed by search engines a ``filter bubble" can be created whereby only supporting information is retrieved \cite{pariser}. As highlighted in the earlier example, this can be problematic in the case of health searching. In this section, we investigate in what ways the user influences the content of the SERP and the medical concern expressed therein. This question can be separated into two aspects: How do SERPs vary for a given query and how do SERPs vary between queries. 

\subsection{Same Query, Different SERPs}
To determine the impact of the user on SERPs within each query, we first investigated how diverse those SERPs were to begin with. We analyzed the composition of SERPs of the 29 most frequent queries from our data set. (Each of these occurred at least 500 times.) We found that on average, 92\% of SERPs contain the same top result. For example, the page ``www.thefreedictionary.com/headache'' appears as the top result for the query ``headache definition'' for almost every user. The three most common top results together covered over 99\% of SERPs. If we consider the Top 8 results on the SERP, we find that eight specific pages are enough to account for 61\% of all results. Hence, most SERPs returned for a given query are highly similar. We do see considerably more diversity when we consider the ordering of pages on the SERP. Hence, we conclude that factors such as time of day, user location and user personalization may shuffle the ordering of pages, especially in the lower half of the SERP, but do not have the power to promote completely different pages to the top the majority of the time (one notable exception to this is personal navigation \cite{teevan}, which agressively promotes pages that an individual visits multiple times, but the coverage of this approach is small). Since user outcomes are driven chiefly by top results, we thus expect the impact of non-query factors on user outcomes to be limited.

Nonetheless, we investigated the impact of prior search activity on the SERP by measuring the association between {\it BeforeSearching} features and {\it SERPConcern} while controlling for query choice. We measure significance as described in section 3.3. We split on {\it SERPConcern} and compared the empirical means of the {\it BeforeSearching} features. Figure 3.1 shows the percentage difference between the mean of the union of the upper buckets and the mean of the union of the lower buckets, aggregated over a 24 hour window before the landmark query (the largest window where significance was obtained). The p-value is shown above the bars. We only show {\it BeforeSearching} features that were significant. 

We see that users who received concerning SERPs relative to other users entering the same query searched for 3\% more serious conditions and 2\% more for the most frequent serious conditions in the 24 hours before the landmark query than than those receiving less concerning SERPs. So, there is an impact of prior searching on the SERP, albeit, as expected, a small one. The difference between the buckets becomes much larger when we consider only users who receive the 10\% most and least concerning SERPs within each query. Figure 3.2 shows that users receiving especially concerning SERPs search for 12\% more serious conditions, search 18\% more for their most frequent serious condition and conduct 10\% more medical searches overall in the 24 hours before the landmark query. If it was the case that prior searching indicates heightened concern, then search engines present significantly more concerning search results to already concerned users, which may be undesirable.

\subsection{Different Queries, Different SERPs}

\begin{figure*}
\centering
\includegraphics[scale=1.8]{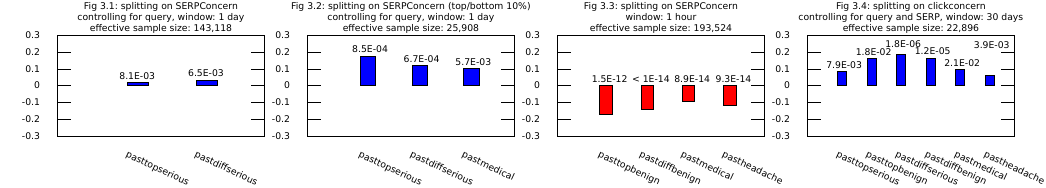}
\includegraphics[scale=1.8]{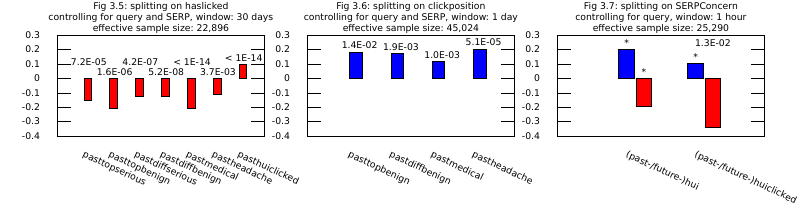}
\caption{Association between user search features and features related to the landmark query. Analysis was conducted as described in section 3.3. In each case, we show the ratio between the search behavior feature value across all upper buckets and across all lower buckets. Hence, positive values imply a positive association between features and vice versa. The p-value is noted next to the bar. We do not show results that were not significant, except in Figure 3.7, were only {\it futurehuiclicked} was significant. The effective sample size refers to the total number of individuals / datapoints included in the analysis. This varies for three reasons. (1) Some features are unavailable for some users. For example, in Figure 3.6, we cannot include users that did not click on the SERP. (2) In some Figures, we only include a specific subset of users. For example, in Figure 3.2, we only include users receiving the 10\% most / least concerning SERPs for each query. (3) We have to subsample for statistical reasons (see section 3.3).}
\end{figure*}

Now, we turn our attention to how SERPs vary across queries. As a starting point, we tested the association between {\it BeforeSearching} and {\it SERPConcern} (without controlling for query choice). Results are shown in Figure 3.3. {\it BeforeSearching} values shown were aggregated over a one hour window. This window size yielded the most significant results, but significance was preserved over all window sizes up to 90 days. Interestingly, users receiving concerning SERPs are now significantly less likely to engage in medical searching before the landmark query. They search for 14\% less benign conditions and 17\% less for their most frequent benign condition. Since these values (and p-values) are much larger than in Figure 3.1, we conclude that the choice of query is the cause for the majority of this effect. At face value, this effect may be either desirable or counterproductive depending on whether {\it BeforeSearching} is an indicator of subjective concern or objective health state.

In Figure 2.1, we break down {\it SERPConcern} by high-level {\it QueryFormulation} features. Error bars are much smaller than those for {\it pastmedical} because the SERP scoring function is not sparse. We see that including an additional symptom in the landmark query (feature: {\it othersymptom}) significantly lowers the level of concern in the SERP. This might be because most of these symptoms are not indicative of a serious brain condition (e.g. cough, stomach pain, hot flashes etc.), thus providing evidence to the search engine that such a condition might not be the underlying reason of the user's health state. We note that this association is the opposite of the association between {\it othersymptom} and {\it pastmedical}. We hypothesize that this effect might be responsible for the negative association in Figure 3.3. Indeed, if we exclude searches that include additional symptoms, all negative associations in Figure 3.3 disappear and we obtain significant positive associations between {\it SERPConcern} and prior serious condition searching. 

In contrast, searching specifically for conditions that explain headache (feature: {\it goal:condition}) yields the most concerning pages. This is logical given what makes a page most concerning is content about (serious) conditions. Again, it is difficult to interpret the meaning of most of the {\it QueryFormulation} features intuitively, but each feature is highly discriminative with respect to {\it SerpConcern} and is thus warranting further study.

Figure 2.2-2.5 breaks down SERPConcern by low-level {\it QueryFormulation} feature. It appears that search engines do a decent job at ranking different symptoms based on how medically serious they are. {\it SERPConcern} is overall much higher amongst relatively concerning symptoms versus relatively benign symptoms, showing that the search engine does respond to this factor. Furthermore, we see a form of consistency in the search engine in that queries with additional symptoms consistently receive below average concern scores. We compared this against different key phrases describing the feature {\it location} (see Table 3) such as ``above eye'', ``temporal'' and ``base of skull''. If we model the {\it SERPConcern} value associated with each additional symptom as well as each location descriptor as normally distributed, then the difference in mean between the two normals is highly significant ($p < 10^{-6}$).

Nonetheless, we still see that there is a considerable amount of unexplained variation in {\it SERPConcern} values. For example, a user entering ``bad headache'' will receive a less concerning SERP than a user entering ``terrible headache''. We hypothesize that it is unlikely that this difference is always due to hidden semantics, but may often be attributable to random noise. This suggests that there is still significant scope for search engines to better reflect the medical content of queries and become more consistent. One caveat is that some of this noise might be caused by our classifier that is used to assign {\it SERPConcern} values. The presence of this noise implies that different user populations which have a preference for certain query formulations may be inadvertently led down completely different page trails to different health outcomes.

\subsection{Summary}

In summary, we find that search engines do seem to show the ability to return less concerning SERPs for benign symptoms in queries as opposed to more serious symptoms. However, there is significant work to be done to achieve a state where search results reflect the medical information given in the query while being robust to irrelevant nuances. Also, we find that past user searching does have a direct impact in making SERPs more concerning, which might not be desirable. Further analysis might discover the exact cause.

\section{RESPONSE OF USER TO SERP}

In this section, we investigate how users respond to the SERP returned in response to the landmark query. We phrase this as two sub-problems: How do users with different levels of prior searching interpret the SERP by making click choices and what impact does the SERP have in altering the behavior of the user. 

\subsection{Impact on Click Behavior}

First, we looked at the impact of {\it BeforeSearching} on click decisions. For this, we measured the association between {\it BeforeSearching} and {\it ClickFeatures}. However, to capture the impact of user predisposition, we can only compare users against each other who not only entered the same query, but also received identical SERPs, to control for those two confounders. Unfortunately, this means we cannot include users in our study that received a SERP none else received, which significantly reduces the size of our effective dataset to lie between 20,000 and 50,000 and users. We split each subpopulation based on high / low values of {\it hasclicked}, {\it clickconcern} and {\it clickposition} and compared the means of {\it BeforeSearching} features. Results are shown in Figures 3.4 (window: 30 days), 3.5 (window: 30 days) and 3.6 (window: 24 hours) respectively. 

We found that users who select more concerning pages on any given SERP are significantly more likely to have conducted medical searches in the 30 days before the landmark query. They have searched for 19\% more serious conditions over this time period. This is quite large given the size of the aggregation window and obfuscating factors such as the limited amount of information available about a page in a SERP caption, the ad hoc nature of a click decision in general and the overall preference for pages ranked near the top. This suggests that people are selectively seeking information and concerned users reinforce their opinion by focusing on concerning content on the SERP. Selective exposure to information has been studied in detail in the psychology community \cite{frey}. The fact that the significance of these results is highest for a large aggregation window illustrates a user's page preference is formed over the medium term, suggesting it is an attitude rather than a momentary state. Furthermore, this result points to past medical search activity as a good proxy for level of health concern, making our previous analyses more meaningful.

We found that users with more prior health searching are more likely to click lower positions on the SERP and are less likely to click overall. This might be because users who have searched about similar topics before are likely to look for specific kinds of information and are thus more likely to reject pages as unsuitable, leading them further down the SERP and ultimately to abandon more of their queries. Connections between topic familiarity and search behavior have been noted in previous work \cite{kelly}. Interestingly, {\it huiclicked} breaks that trend and is positively associated with clicking on the SERP. This might be because {\it huiclicked} is by definition associated with a user's general disposition to click on a SERP, which in turn affects the probability of the user clicking in response to the landmark query. 

\subsection{Impact on Future Behavior}

Now we turn to measuring the impact that the SERP has on the state of the user as measured by {\it AfterSearching}. Previous research (e.g., survey responses in \cite{whitehorvitzcyber,foxduggan}) showed that online content can have a direct impact on people's healthcare utilization decisions. It also has been shown that the content of pages viewed can be used to predict future serious condition searches \cite{whitehorvitzesc}. We believe that finding evidence in logs that different SERPs cause users to respond differently would strengthen the motivation for improving search engine performance during health searches. 

Before proceeding, we must point out that this task is quite difficult. We have already shown that users have a myriad of significant predisposition which shape the SERP. Hence, it is impossible to tell whether any change in user behavior after the landmark query was really caused by the SERP or is the result of a predisposition.

One way to mitigate this is, again, control for query choice. Unfortunately, the group of users receiving concerning SERPs within each query is very different from the group of users receiving concerning SERPs overall, as within-query variation of {\it SERPConcern} is much smaller than between-query variation (see Section 5). Hence, the true impact of the SERP on users is likely much larger than is measurable in this experiment.

Splitting on {\it SerpConcern} and comparing the means of {\it AfterSearching} features yielded no significant differences over any aggregation window. However, looking at the top 10\% vs. bottom 10\% of SERPs within each query yielded surprising results. They are shown in Figure 3.7 (window: 1 hour). Users receiving especially concerning SERPs perform 20\% less HUI queries in the hour following the landmark query when compared to their peers who entered the same query but received an especially unconcerning SERP. Additionally, these users had a larger empirical probability of expressing HUI in the preceding hour (non-significant), amplifying the drop. The ratio of HUI queries between users receiving concerning vs. non-concerning SERPs changes from +20\% to -20\% from past to future. To solidify this result, we reran our experiment under exclusion of users who had performed any medical searching over the last 1 hour, 24 hours, etc. We found that the negative association of {\it futurehui} and {\it futurehuiclicked} with {\it SERPConcern} remained significant even when excluding users who had performed no medical searching in the preceding 30 days. (Note that the more users we exclude, the less data we have and the more difficult it is to achieve significance.) This is a surprising result that warrants further investigation.

Even though there is no causal argument to be made, we were still interested in the outright association of {\it SERPConcern} and {\it AfterSearching}. We made the interesting observation that the association of {\it topserious} and {\it diffserous} actually decreased from past to future. For example, users receiving concerning SERPs issued 2\% more searches for their most frequent serious condition during the hour before the landmark query, but 1\% less afterwards (both non-significant). On the contrary, users receiving concerning SERPs searched 17\% less for their most frequent benign conditions in the hour before the landmark query and only 8\% less in the hour after. These results are difficult to interpret. However, they do provide some evidence that the impact of the SERP on the user's concern might be more subtle than expected.

\subsection{Summary}
In summary, we found that concerned users are significantly more likely to click concerning pages which may be a serious problem. Additionally, we found only weak evidence that concerning SERPs cause an increase in user's health concerns. To the contrary, users receiving concerning SERPs may be less likely to seek HUI in the short term. Both of these effects need to be investigated further.

\section{DISCUSSION AND IMPLICATIONS}

In this section, we summarize our findings and discuss implications, limitations and opportunities for future research. Our main findings are as follows:

\begin{itemize*}
\item Innocuous details in query formulation can hold characterizing information about the search engine user.
\item While the search engine does respond to general trends in query formulation sensibly, there is a lot of variation in SERPs for different but similar word choices. 
\item While users with a history of medical searching are significantly more likely to pick out especially concerning content, the search engine also serves those users more concerning SERPs to begin with.
\item There was only weak evidence for the effect of intensification of health concerns through concerning SERPs, but some evidence that concerning SERPs might reduce HUI queries and hence real-world health seeking in the short term.
\end{itemize*}

Considering this, we believe that the most important implication for improving medical search results from this study is clearly highlighting the need for a further rigorous quantification of the extent to which concern in SERPs influences users' levels of health concern. Utlimately, through methodologies to study consenting user cohorts in detail, we can move beyond our focus on health concerns to target health anxieties directly.

We also believe there is significant  potential in refining our understanding of the meaning of both landmark query formulation and general search behavior for extracting hidden information about the user. In this paper, we have shown how these might be used to better infer levels of health concern. In this way, we might be able to improve search engine performance, for example by promoting more medically trusted pages that discuss a wide variety of possible causes objectively if we suspect health anxiety in the searcher. However, a lot more can be done. For example, word choices might reveal the age of the user \cite{torres} or their level of domain expertise \cite{hembrooke}, which in turn would have implications for the medical meaning of symptoms. We also have not yet considered formulation nuances of queries other than the landmark query, which might hold much richer information. The need for such analysis to adjust for anxiety is exemplified by our finding that users who conduct more medical searching have a preference for concerning content, which might indicate a cycle of self-reinforcing concern.

Other questions for future research include: In how far is frequent medical searching indicative of the user's actual state of health vs. subjective concern? How can we effectively integrate the results of a possible algorithm that learns searchers' true medical situation into a standard retrieval model without ``overmanaging'' or hurting the overall robustness of the engine? How is it possible to find the right balance between preventing health anxiety and the potential delay of important medical treatment when users are confronted with only benign explanations when they are actually sick?

Although we studied the behavior many searchers, one  limitation is that we used a single search engine, meaning our findings pertaining to user behavior might not generalize. Another limitation is the imperfect measurement of {\it SERPConcern}. Due to the fact that some queries occur very frequently (e.g. ``headache''), any error incurred by our scoring function on the top results for those queries might have a big impact on the outcome. As mentioned earlier, another limitation is that we only used data on queries that were related to headache. Finally, query logs offer only a limited view on health concerns, and we need to work with searchers directly to understand the motivations behind their search behaviors.

\section{CONCLUSIONS}

We presented a log-based longitudinal study of medical search behavior on the web. We investigated the use of medical query formulation as a lens onto searchers health concerns and found that those features were predictive for this task. We evaluated how a major search engine responds to changes in medical query formulation and saw that there were some trends and a lot of variance, which might have adverse affects by way of misinformation. We showed that a significant tendency for medically concerned users to view concerning content makes it important for engines to manage this effect (e.g., by considering estimated level of health concern as a personalization feature). Finally, we raised the need for detailed study of the impact of SERP composition on users' future behavior. We believe that our results can function as an initial guide for developing practical tools for better health search as well as to inform deeper investigations into health concerns and anxieties on the Web and in general.


\bibliographystyle{abbrv}

\bibliography{sigproc}

\end{document}